\documentclass{icrc29}
\usepackage{graphicx} 
\newcommand{\eqb}{\begin{eqnarray}}
\newcommand{\eqe}{\end{eqnarray}}
\newcommand{\diff}{{\rm d}}
\newcommand{\gammapeak}{\gamma_{\rm p}}
\newcommand{\gammamax}{\gamma_{\rm max}} 
\newcommand{\gammamin}{\gamma_{\rm min}}
\newcommand{\gammawind}{\gamma_{\rm w}}

\newcommand{\mnras}{MNRAS}
\newcommand{\apj}{ApJ}
\newcommand{\aap}{A\&A}
\newcommand{\apss}{Astrophys.\ Space Sci.}

\begin{document}

\title[Gamma-rays from PSR~B1259~$-$63]{Modelling the gamma-ray 
emission from PSR~B1259~$-$63}
\author[Kirk, Ball \& Johnston]{J.G. Kirk$^a$, Lewis Ball$^b$,
  S. Johnston$^c$%
\\
(a) Max-Planck-Institut f\"ur Kernphysik, D-69029 Heidelberg, Germany\\
(b) Australia Telescope National Facility, CSIRO, PO Box 276, Parkes, NSW 2870,
Australia\\
(c) Australia Telescope National Facility, CSIRO, PO Box 76, 
Epping, NSW 1710, Australia}
\presenter{Presenter: J.G. Kirk ( John.Kirk@mpi-hd.mpg.de ), \
ger-kirk-J-abs1-og22-oral}
\maketitle
\begin{abstract}
The high-energy gamma-ray emission discovered using the H.E.S.S. telescopes 
from the binary system
PSR~B1259~$-$63, is modelled using an extension of 
the approach 
that successfully 
predicted it.  
We find that the 
simultaneous INTEGRAL and H.E.S.S. data permit both a model with
dominant radiative losses, high pulsar wind Lorentz factor and 
modest efficiency as well as one with dominant adiabatic losses, a slower wind
and higher efficiency. Additional, simultaneous, X-ray and TeV 
data sets are needed to lift this degeneracy. 
\end{abstract}
\section{Introduction}
The radio pulsar PSR~B1259~$-$63 is in a highly
elliptical orbit about the luminous Be~star SS2883.
Pulsar winds 
are expected to accelerate electrons to
Lorentz factors of up to $10^7$, leading to up-scattering of 
the ultra-violet
photons from the Be~star into the TeV range 
\cite{tavaniarons97,kirkballskjaeraasen99,chernyakovaillarionov00,kawachietal04}.
This can happen before or after the wind passes through 
its termination shock \cite{ballkirk00}. However, 
in the case of PSR~B1259~$-$63, 
the time spent by an individual electron in the unshocked wind is 
short compared to the time spent in the vicinity of the Be star after 
passing the shock. Thus, even if the shock simply isotropises the
electrons without energising them, the post-shock emission
should dominate over the pre-shock emission. This conclusion holds
a forteriori, if, as expected, the shock transfers some of the  
incoming kinetic energy into nonthermal particles.

Observations using the H.E.S.S. array of imaging \v{C}erenkov telescopes 
around and after the periastron
passage in early 2004 detected a strong signal in the TeV range
\cite{schlenkeretal05,aharonianetal05}. The measured spectrum is
in excellent agreement in both slope and absolute normalisation 
with that predicted by a model in which the post-shock pulsar wind
electrons have a simple, single power-law distribution 
\cite{kirkballskjaeraasen99}. Significant
night-to-night fluctuations in the TeV light curve as well as an overall
decrease on the timescale of months were also 
observed by H.E.S.S., possibly
correlated with variations in the unpulsed radio emission
\cite{johnstonetal05}.  
Short timescale fluctuations, especially close to periastron, can plausibly be
attributed to departures from spherical symmetry in the structure of the
pulsar wind or the Be~star wind; the most detailed current model ascribes the 
variation in the unpulsed
radio emission to the latter 
\cite{balletal99,connorsetal02}. These were not taken into 
account in the predictions of the TeV emission
\cite{kirkballskjaeraasen99}, which included only those effects arising from
the variation of the stellar separation over the orbit.  
A model in which the TeV emission
arises from proton-proton interactions 
in the anisotropic wind (\lq\lq disk\rq\rq) 
of the Be~star \cite{kawachietal04}
produces short timescale features 
qualitatively similar to those observed, but appears to  
predict a flatter TeV spectrum than that seen by H.E.S.S. 

In this paper, we present preliminary results from an
extended version of the model of \cite{kirkballskjaeraasen99}. 
Injection of a double power-law electron spectrum, 
similar to that thought to be
injected into the Crab Nebula by its central pulsar
\cite{gallantetal02} is included, as is 
the transition from radiative to 
adiabatic loss mechanisms as the stellar separation increases. 

\section{The model}
It has recently become clear 
\cite{lyubarsky03b}
that the pulsar wind that fuels the Crab Nebula
injects into it relativistic 
electrons and positrons (and possibly ions) whose energy
distribution can be approximated as a double power-law:
$Q(\gamma)=\left(\gamma/\gammapeak\right)^{-q_1}$, for 
$\gammamin<\gamma<\gammapeak$ and 
$Q(\gamma)=\left(\gamma/\gammapeak\right)^{-q_2}$ for 
$\gammapeak<\gamma<\gammamax$.
The high-energy index is determined by the slope of the X-ray
spectrum of the Crab Nebula: $q_2\approx 2.2$, in agreement with theories of
first-order Fermi acceleration at relativistic 
shocks \cite{kirketal00,achterbergetal01}. The low energy index follows from
the slope of the radio to infra-red spectrum: $q_1\approx1.6$.  With these
values, most particles are concentrated around the lower cut-off 
at $\gamma=\gammamin$, whereas most of the energy is injected in electrons of
Lorentz factor $\gamma\sim\gammapeak$. In the Crab, $\gammamin\approx100$,
$\gammapeak\approx10^6$ and $\gammamax\approx 10^9$. The resulting synchrotron
spectrum contains two breaks, one due to cooling and one intrinsic to the
injected spectrum (at $10^{13}\,$Hz and $10^{15}\,$Hz in the Crab), 
as well as upper and lower cut-offs. If this injection spectrum is produced at
the termination shock front, and if the cold upstream flow is dominated
by the kinetic energy flux in electron-positron pairs, 
then the Lorentz factor of the wind is 
$\gammawind=\int\diff\gamma\, \gamma Q(\gamma)/\int\diff\gamma\, Q(\gamma)$.
In the following we adopt this injection model.

In PSR~B1259~$-$63, 
relativistic electrons and positrons in the shocked pulsar wind
suffer adiabatic losses as the plasma expands away from the shock
front, as well as radiative losses by synchrotron radiation and inverse Compton
scatterings, primarily of the ultra-violet photons from the Be~star. The
energy dependence of these loss processes is different and influences the
resulting distribution function. Two sets of models were  constructed in 
\cite{kirkballskjaeraasen99}: one for dominant
adiabatic and one for dominant radiative losses. Both 
were calibrated using the observed X-ray synchrotron emission, and provided 
accurate
predictions of the TeV spectrum subsequently detected 
just before periastron. However, the two models imply quite different injection
spectra. 

As the pulsar moves away from the Be~star, both the target radiation field and
the magnetic field where the winds interact decrease, along with the gas
pressure. For the toroidal field structure 
expected in a pulsar wind, the ratio of the energy densities of 
magnetic field and target radiation remain constant, so that, in the 
absence of Klein-Nishina effects, the ratio of synchrotron to inverse Compton
radiation should not vary with binary phase. However, if the expansion time
scales linearly with the stellar separation, adiabatic losses become
more important with respect to radiative losses as the stars move apart.
In the models used here, we account for this by switching between a radiative 
and an adiabatic loss term 
in the kinetic equation of the electrons at the Lorentz factor where the loss
rates coincide. The losses themselves are specified by the magnetic field
strength in the emission region, and the adiabatic loss time scale at
periastron. 

\begin{table}
\caption{%
\label{parameters}%
The model parameters. The efficiency refers to the fraction of the
  spin-down luminosity injected into the source as relativistic particles
  (assuming a source distance of $1.5\,$kpc). The adiabatic loss time 
$t_{\rm ad}$ is given in units of the light crossing time of the periastron
  separation ($320\,$sec). $B$ is the 
magnetic field strength in the source at periastron %
}
\begin{tabular}{l|lllllll}
Model:&$\gammamin$&
$\gammapeak$&$\gammamax$&$\gammawind$&$B$&Efficiency&$t_{\rm ad}$\\
\hline         
&&\cr
A &$425$ &$10^7$  &  $5\times10^7$ &$5.5\times10^4$ &$0.3\,$G &$10$\% &15  \\
B &$425$ &$10^6$ &$4\times10^7$ &$2.9\times10^4$  &$0.3\,$G &$100$\% & 0.5\\
\end{tabular}
\end{table}
 
The emitted radiation is a combination of synchrotron radiation in a uniform
magnetic field and inverse Compton scattering of ultra-violet photons from the
Be~star. On its way from the pulsar system to the observer the inverse-Compton
emission is partially reabsorbed via pair production on the stellar photons
\cite{kirkballskjaeraasen99}. These processes are well-understood. 
To expedite the computations, \cite{kirkballskjaeraasen99} used
delta-function approximations to the emissivities of both emission
mechanisms. In addition to the standard assumption (of synchrotron theory)
that the direction of the emitted photons is approximately that of the
incoming electron, these
approximations replace the energy spread of 
photons emitted by a given electron by an appropriate monochromatic
term. On physical grounds, this approximation can be expected to
give better results 
as the electron Lorentz factor increases and to be especially 
accurate in the Klein-Nishina regime of inverse Compton scattering. 
We have checked these approximations using a computationally
more costly evaluation of the full Klein-Nishina rates and 
find they indeed give very accurate results, in conflict with
the comparison presented in \cite{khangoulianaharonian05}. 
On the other hand, care must be taken with
these approximations for synchrotron radiation, 
especially when sharp gradients in
the distribution function are present. This is because the 
synchrotron process is 
equivalent to a scattering
deep inside the Thomson regime, leading to a relatively broad emission
cone. In this paper we use the full synchrotron emissivity, but keep the
delta function approximation for inverse Compton scattering used in
\cite{kirkballskjaeraasen99}. 

\section{Results}
Modelling the spectrum and light curve of the high energy emission during
the 2004 periastron passage is made difficult by the scarcity of simultaneous 
TeV and X-ray data sets. The only ones available
are the X-ray/soft gamma-ray spectrum detected by INTEGRAL 
between 14 and 17 days after
periastron passage \cite{shawetal04} and the March
2004 observations by H.E.S.S. \cite{aharonianetal05}.
These data are not sufficient to determine the dominant loss mechanism.
To illustrate this, we consider two models, with parameters given in
Table~\ref{parameters},
where we also quote the efficiency of each model,
defined as the ratio of the power injected into the emission 
region in the form of relativistic electrons to the pulsar spin-down power.

\begin{figure}
\hbox{%
\includegraphics[width=0.25\textwidth]{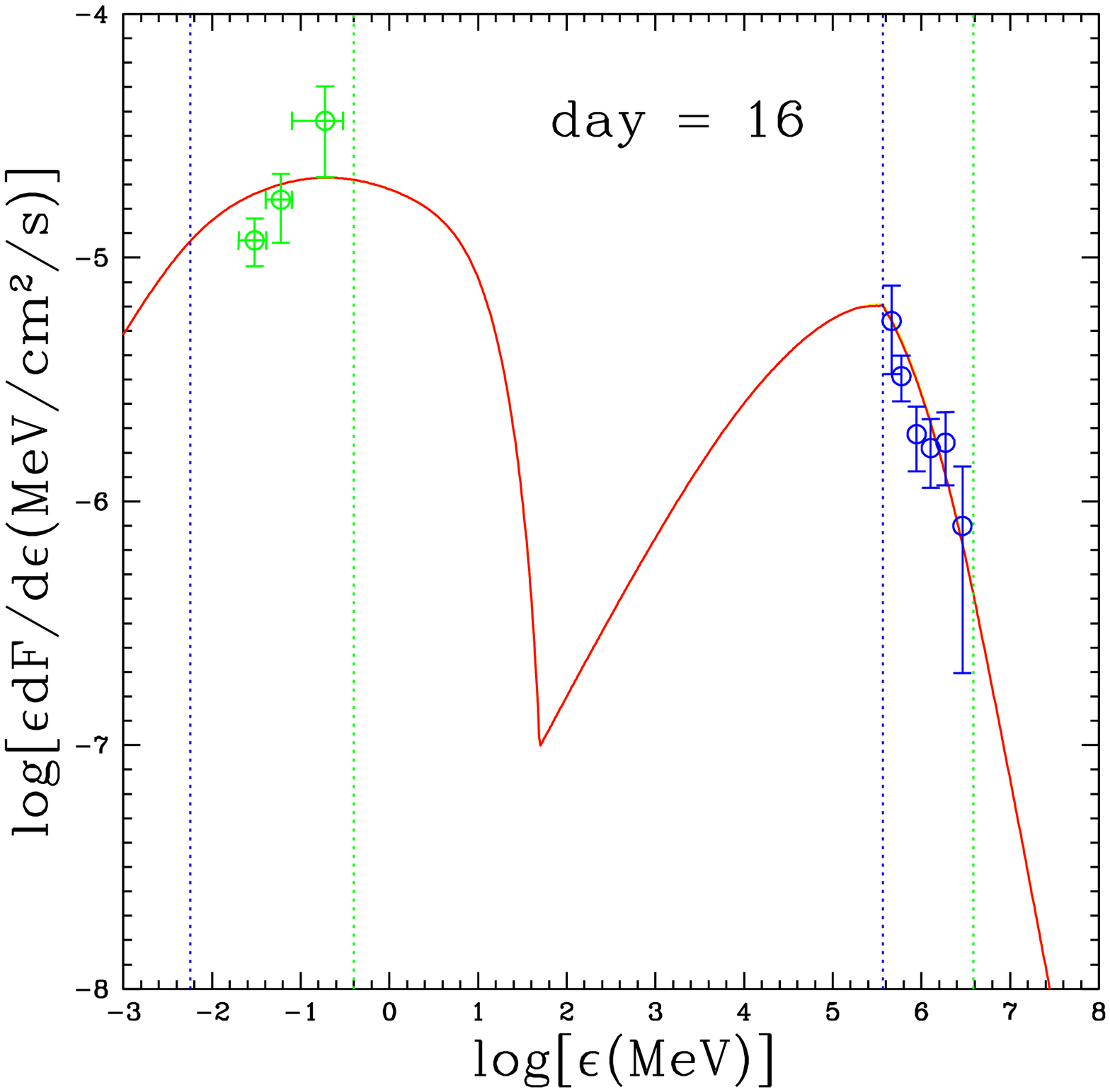}%
\includegraphics[width=0.25\textwidth]{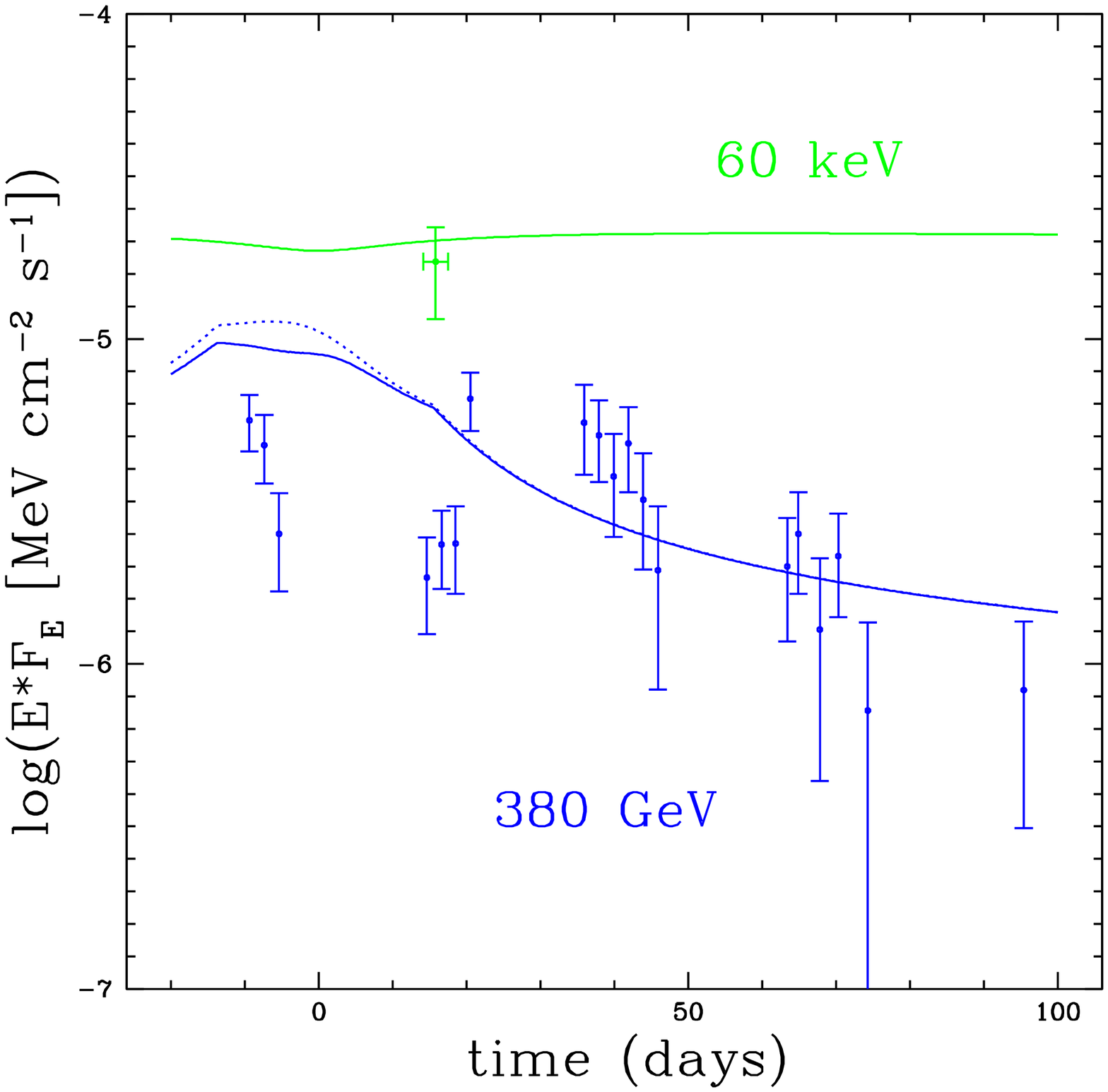}%
\includegraphics[width=0.25\textwidth]{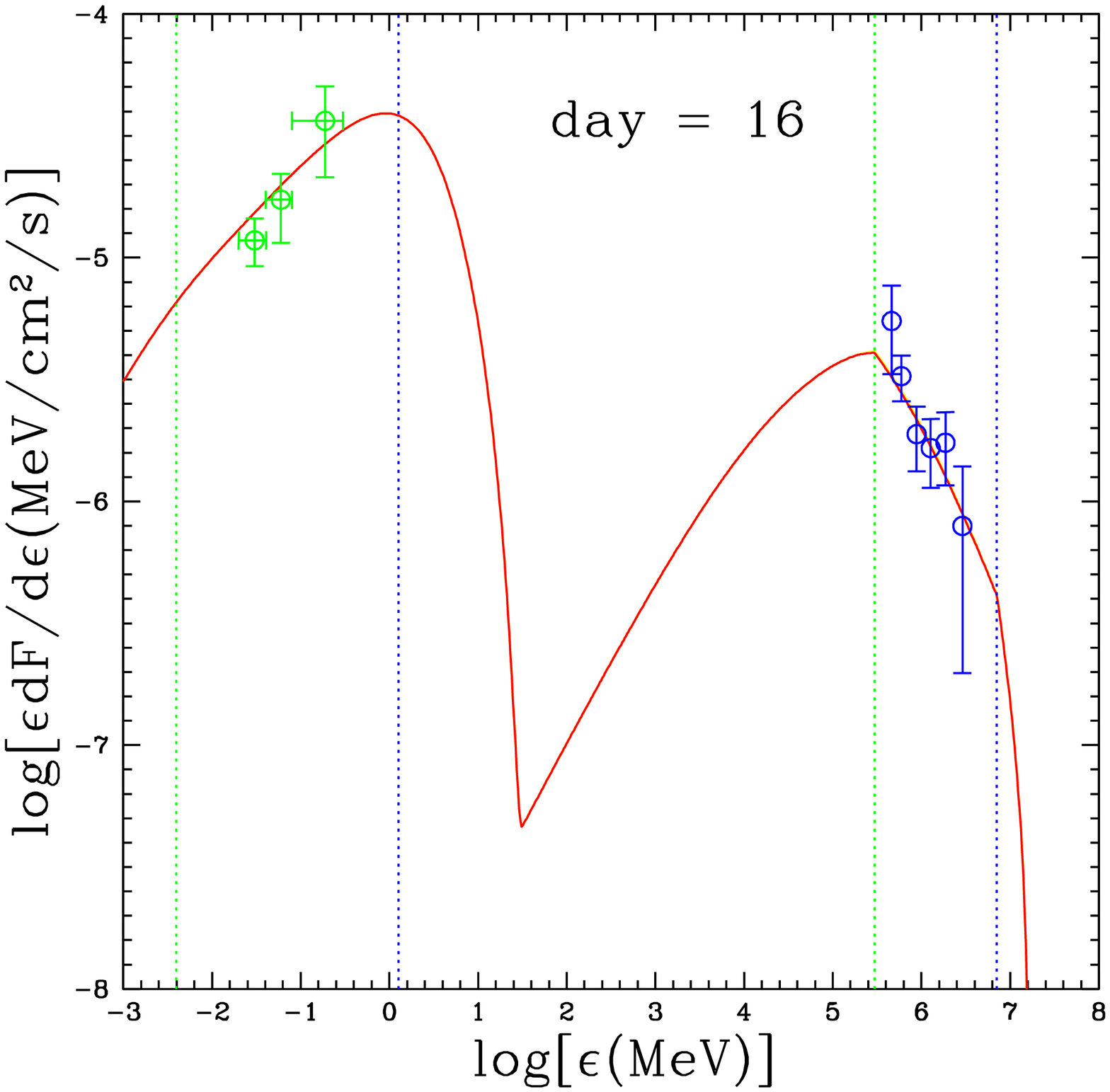}%
\includegraphics[width=0.25\textwidth]{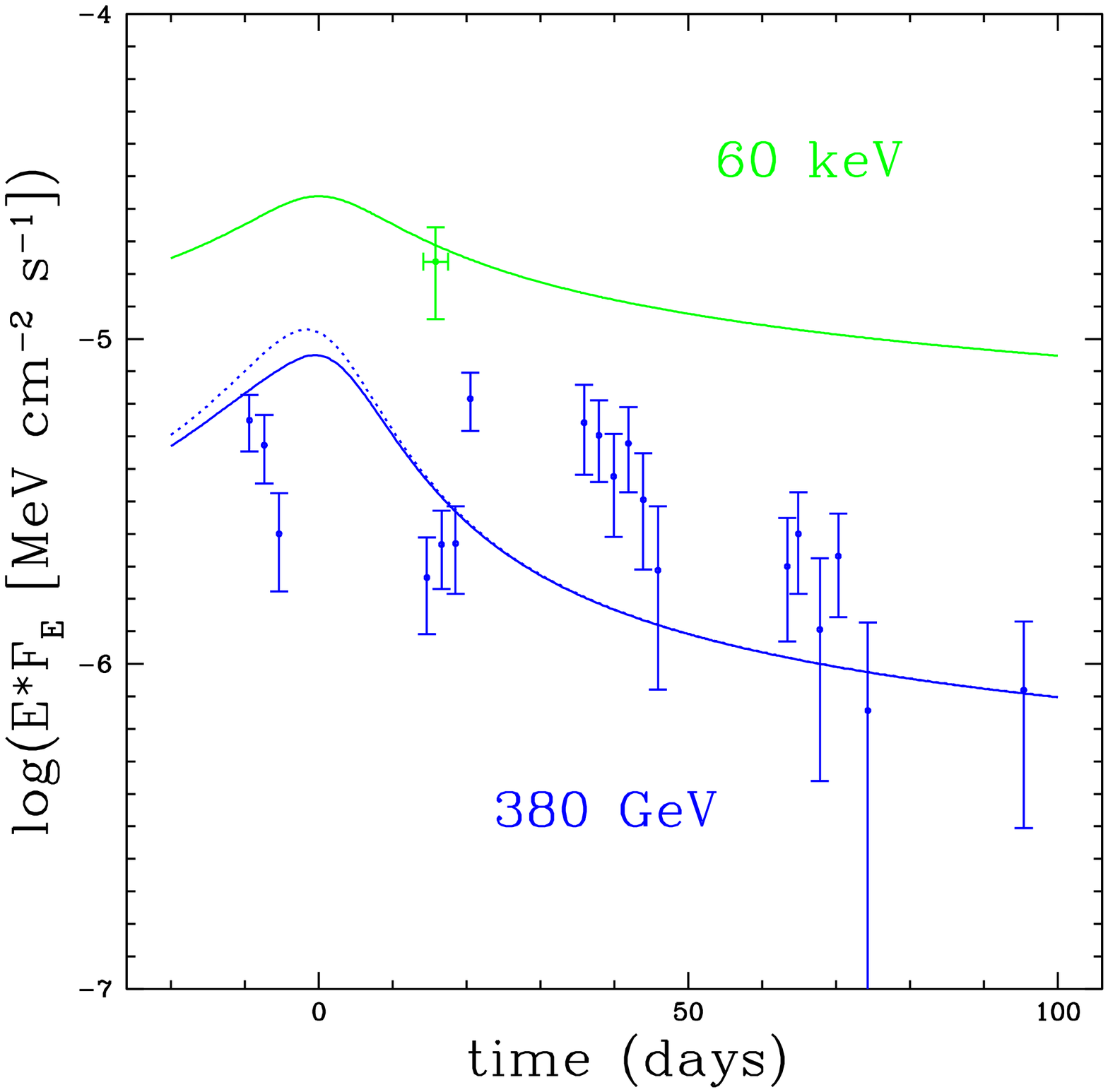}%
}
\caption{%
\label{spectrum}%
Spectra and light curves of models~A (left) and B 
(right). At periastron, radiative (adiabatic) losses dominate in model~A~(B).
The change in character of the 
$380\,$GeV light curve in Model~A 
17 days after periastron arises because at this epoch the corresponding  
electrons pass into the adiabatic loss-dominated regime.  
}
\end{figure}

The TeV gamma-ray spectrum implies that the differential 
number of radiating
electrons is roughly $\diff\log N/\diff\log\gamma\approx -2.5$.
This can be modelled either as the result of radiative 
cooling of the hard injection spectrum at
$\gamma<\gammapeak$, or of the accumulation without energy loss of 
electrons injected at $\gamma>\gammapeak$. The first case applies if
radiative losses
dominate. The second, if the losses are adiabatic.

Spectra and light-curves for Model~A 
(radiative-loss dominated; left panels) and 
Model~B (adiabatic-loss dominated; right panels) together with 
observations by INTEGRAL
(green) \cite{shawetal04} and H.E.S.S. (blue) \cite{aharonianetal05}
are shown in Fig.~\ref{spectrum}.
The vertical blue lines depict the photon energies associated with electron
Lorentz factors at which the adiabatic and radiative cooling rates are equal. 
At these points, a \lq\lq cooling
break\rq\rq\ appears in the spectrum. The vertical green lines mark the break
intrinsic to the injection function, at $\gamma=\gammapeak$.
The dotted curve indicates the intrinsic emission, before propagation
through the radiation field of the Be~star.

It is evident that Model~A, 
cannot produce as good a fit to the (hard)
X-ray spectrum as Model~B, in which adiabatic losses dominate.
The reason is that, in the hard X-ray range, cooling by synchrotron emission
is more important than cooling by inverse Compton emission. In this case, the 
hardest possible model spectrum produced by cooled electrons 
has a photon index of 
$-1.5$. This limitation does not apply to the adiabatic-loss dominated models. 
However, the data are not sufficient to reject Model~A, 
because of the relatively large error associated with the 
spectrum reported by INTEGRAL ($-1.3\pm.5$). It should be noted that 
in {\em soft} X-rays, radiative-loss 
dominated models produce a harder spectrum. 
This interesting effect, (remarked upon by many authors including
\cite{kirkballskjaeraasen99}) arises from the transition between inverse
Compton cooling at low frequencies and synchrotron cooling at higher
frequencies as the scattering regime moves from Thomson into Klein-Nishina.   

The key property of the models --- that
the TeV spectrum is formed by electrons of $\gamma<\gammapeak$ in Model~A
(radiation losses) and $\gamma>\gammapeak$ in Model~B (adiabatic
losses) --- implies a faster pulsar wind in Model~B, but also a very high
efficiency. Since the computed efficiency 
depends on the poorly known distance to
this object, values in excess of $100\%$ are not forbidden. Nevertheless, 
they are uncomfortable, since the spin-down power must also 
provide for the kinetic energy of the bulk post-shock 
flow and the Poynting flux, as well as, perhaps, relativistic ions.
In Model~B, this problem is exacerbated by the 
extremely short 
($160\,$secs) adiabatic loss time, which implies a high post-shock
bulk speed.


\end{document}